\documentclass[twocolumn,showpacs,preprintnumbers]{revtex4}%
\usepackage{amssymb}
\usepackage{amsmath}
\usepackage{graphicx}
\usepackage{dcolumn}
\usepackage{bm}
\usepackage{amsfonts}%
\begin{document}
\title{On-Fiber Optomechanical Cavity}
\author{Ilya Baskin}
\affiliation{Department of Electrical Engineering, Technion, Haifa 32000 Israel}
\author{D.Yuvaraj}
\affiliation{Department of Electrical Engineering, Technion, Haifa 32000 Israel}
\author{Gil Bachar}
\affiliation{Department of Electrical Engineering, Technion, Haifa 32000 Israel}
\author{Keren Shlomi}
\affiliation{Department of Electrical Engineering, Technion, Haifa 32000 Israel}
\author{Oleg Shtempluck}
\affiliation{Department of Electrical Engineering, Technion, Haifa 32000 Israel}
\author{Eyal Buks}
\affiliation{Department of Electrical Engineering, Technion, Haifa 32000 Israel}
\date{\today }

\begin{abstract}
A fully on-fiber optomechanical cavity is fabricated by patterning a suspended
metallic mirror on the tip of an optical fiber. Optically induced self-excited
oscillations of the suspended mirror are experimentally demonstrated. We
discuss the feasibility of employing on-fiber optomechanical cavities for
sensing applications. A theoretical analysis evaluates the sensitivity of the
proposed sensor, which is assumed to operate in the region of self-excited
oscillations, and the results are compared with the experimental data.
Moreover, the sensitivity that is obtained in the region of self-excited
oscillations is theoretically compared with the sensitivity that is achievable
when forced oscillations are driven by applying an oscillatory external force.

\end{abstract}
\maketitle





Optomechanical cavities, which are formed by coupling an optical cavity with a
mechanical resonator, are currently a subject of intense
study.\cite{Marquardt_0905_0566, girvin_Trend,Kippenberg_1172} Such systems
may allow experimental study of the crossover from classical to quantum
mechanics (see Ref. \cite{Meystre_1210_3619} for a recent review). In
addition, optomechanical cavities can be employed in optical communications
\cite{Wu_et_al_06} and in other photonics
applications.\cite{Stokes_et_al_90,Hossein-Zadeh&Vahala_10} When the finesse
of the optical cavity that is employed for constructing the optomechanical
cavity is sufficiently high, the coupling to the mechanical resonator that
serves as a vibrating mirror is typically dominated by the effect of radiation
pressure.\cite{Kippenberg_1172} On the other hand, bolometric effects can
contribute to the optomechanical coupling when optical absorption by the
vibrating mirror is significant.\cite{Metzger_1002, Jourdan_et_al_08,
Metzger_133903, Marino&Marin_10} In general, bolometric effects play an
important role in relatively large mirrors, in which the thermal relaxation
rate is comparable to the mechanical resonance frequency.
\cite{Aubin_et_al_04, Marquardt_103901, Paternostro_et_al_06,
Liberato_et_al_10} Phenomena such as mode cooling and self-excited
oscillations
\cite{Hane_179,Kim_1454225,Aubin_1018,Carmon_223902,Marquardt_103901,Corbitt_021802,Carmon_123901,Metzger_133903}
have been shown in systems in which bolometric effects are dominant.
\cite{Metzger_133903, Metzger_1002, Aubin_et_al_04,
Jourdan_et_al_08,Zaitsev_046605,Zaitsev_1589}

In this paper we study a novel configuration, in which an optomechanical
cavity is constructed using a single mode optical fiber. In this configuration
the vibrating mirror is fabricated on the tip of an optical fiber. Additional
static reflector is introduced in the fiber, so that optomechanical cavity is
formed. We experimentally demonstrate that self-excited oscillations of the
vibrating mirror can be induced by injecting a monochromatic laser light into
the fiber. This effect is attributed to the bolometric optomechanical coupling
between the optical mode and the mechanical resonator.

The ability to optically induce self-excited oscillations can be exploited for
operating an on-fiber optomechanical cavity as a sensor. Such a device can
sense physical parameters (e.g. absorbed mass, heating by external radiation,
acceleration, etc.) that affect the resonance frequency of the suspended
mirror. The simplicity of fabrication and operation, the small size and
robustness of such a device and the unneeded actuator coupling and circuitry
might be beneficial for future sensor applications. Further we present the
on-fiber optomechanical cavity showing self-excited oscillations, give a
theoretical estimate of its sensitivity and compare it with the measured performance.

\begin{figure}
[ptb]
\begin{center}
\includegraphics[
height=8.2cm,
width=8.1121cm
]%
{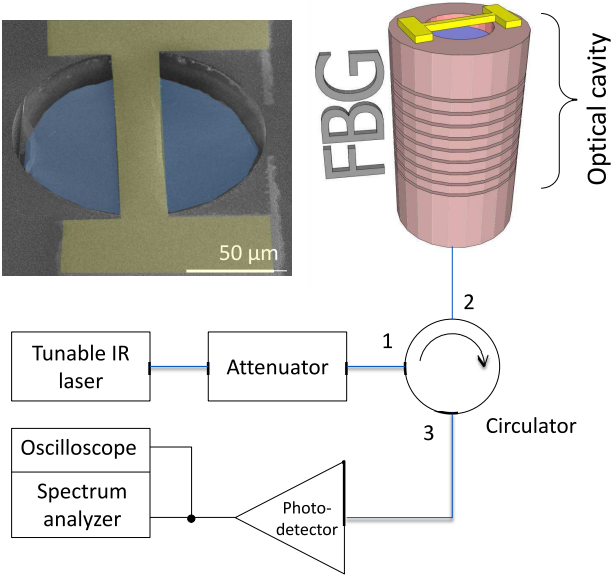}%
\caption{The experimental set-up. On- fiber optomechanical cavity excited by a
tunable laser. The reflected light intensity is measured and analyzed. Inset:
electron micrograph of a suspended micromechanical mirror (color code:
blue-silica fiber, yellow - gold mirror, gray-zirconia ferrule),view is tilted
by 52$^{o}$ .}%
\label{FigExp}%
\end{center}
\end{figure}

The optomechanical cavity seen in Fig. \ref{FigExp} was fabricated on the tip
of a single mode fused silica optical fiber (Corning SMF-28 operating at
wavelength $\lambda\approx1550%
\operatorname{nm}%
$). The processing started from thermal evaporation of chromium and then gold
layers with thicknesses of $10$ and $200%
\operatorname{nm}%
$ respectively, on a flat polished tip of a fiber, held in a zirconia ferrule.
The evaporated layers were directly patterned by a focused ion beam to the
desired mirror shape ($20%
\operatorname{\mu m}%
$ wide doubly clamped beam). Finally, the gold mirror was released by etching
the underlying silica in 7\% HF acid (90 $%
\operatorname{min}%
$ etch time), while the beam remained supported by the zirconia ferrule. The
precise alignment of the micro-mechanical mirror and fiber core that was
achieved in this process allowed robust and simple operation of the fiber-tip
devices without a need of any post fabrication positioning.

The static mirror of the optomechanical cavity was provided by a fiber Bragg
grating (FBG) mirror of high reflectivity (the FBG stopband of $0.4%
\operatorname{nm}%
$ FWHM was centered at $\lambda$) that is made using the phase mask technique
(see Fig. \ref{FigExp}). The length of the optical cavity was $l\approx10%
\operatorname{mm}%
$ providing a free spectral range $\Delta\lambda=\lambda^{2}/2n_{\mathrm{eff}%
}l\approx80$ pm (where $n_{\mathrm{eff}}$ $=1.47$ is the effective refraction
index for SMF-28); five resonant wavelengths were located within the range of
the FBG stopband.%

Monochromatic light was injected into the fiber bearing the cavity on its tip
from a laser source with adjustable output wavelength (between $1525$ and
$1575%
\operatorname{nm}%
$) and power level (up to $20%
\operatorname{mW}%
$). The source was connected through an optical circulator, that allowed the
reflected light intensity to be measured by a fast responding photodetector.
The detected signal was analyzed by an oscilloscope and a spectrum analyzer.
The experiments were performed with samples in vacuum (at residual pressure
below $10$ mPa) thermally anchored to a cold finger with temperature
adjustable between $4$ to $300$ $%
\operatorname{K}%
$. The resonance frequency of the suspended mirror oscillations, $\omega_{0}$
was estimated by the frequency of thermal oscillations measured at input laser
power well below the self-excited oscillations threshold.

\begin{figure}
[ptb]
\begin{center}
\includegraphics[
height=4.6415cm,
width=9.1028cm
]%
{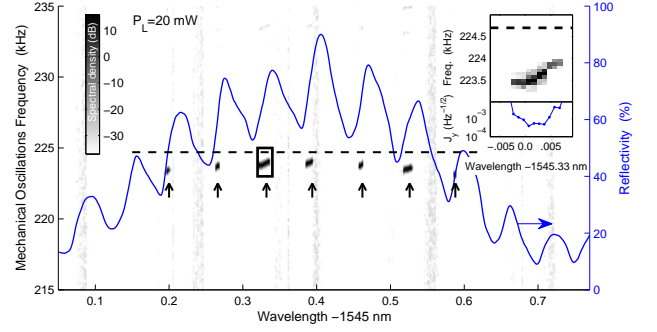}%
\caption{The self-excited oscillations visible as sharp peaks (dark gray
regions on colormap, marked by arrows) in reflected power spectrum are
obtained at optical excitation wavelengths corresponding to positive slope of
the sample reflectivity (shown by a solid curve). The self-excited
oscillations frequency is slightly below the estimated mechanical resonance
frequency, indicated by a dashed line. Zoom-in to a self-excited oscillations
peak (marked by a rectangle) and estimation of sensitivity factor
$J_{\mathrm{y}}$\ are shown in the inset.}%
\label{Fig_osc}%
\end{center}
\end{figure}

In Fig. \ref{Fig_osc} we present the spectral power density of the reflected
light intensity as a function of optical excitation wavelength. The sharp
spectral peaks that appear just below the mechanical resonance frequency
$\omega_{0}$ (which is marked by a dashed line in Fig. \ref{Fig_osc}) indicate
the appearance of the self-excited oscillations of the vibrating mirror. These
oscillations are "turned on" when the incoming laser power exceeds a threshold
value and the laser wavelength is adjusted to the regions of high (positive)
slope of the cavity reflectance spectrum (Fig. \ref{Fig_osc},\ solid curve).
The sign of the reflectivity slope suggests that the beam is
buckled.\cite{Yuvaraj_1207_0947}%

To estimate the sensitivity of such a sensor that is based on optomechanical
cavity the phase noise of the self-excited oscillations is theoretically
evaluated below, and the results are compared with experimental data. In
general, the resonant detection with a mechanical resonator is a widely
employed technique in a variety of
applications.\cite{Larsen_121901,Ilic_et_al_04,Cleland_235,Pandey_et_al_10} A
detector belonging to this class typically consists of a mechanical resonator,
which is characterized by an angular resonance frequency $\Omega$ and
characteristic damping rate $\Gamma$. Detection is achieved by coupling the
measured physical parameter of interest, denoted as $p$, to the resonator in
such a way that $\Omega$ becomes effectively $p$ dependent, i.e.
$\Omega=\Omega\left(  p\right)  $. The sensitivity of the detection scheme
that is employed for monitoring the parameter of interest $p$ can be
characterized by the minimum detectable change in $p$, denoted as $\delta p$.
For small changes, $\delta p$ is related to the normalized minimum detectable
change in the frequency $\sigma_{\mathrm{y}}=\delta\Omega/\Omega$ by the
relation $\delta p=\left\vert \partial\Omega/\partial p\right\vert ^{-1}%
\Omega\sigma_{\mathrm{y}}$. The dimensionless parameter $\sigma_{\mathrm{y}}$,
in turn, typically depends on the noise in the system and on the averaging
time $\tau$ that is employed for the measurement.

A commonly employed detection scheme is based on externally driving the
resonator with a monochromatic force at a frequency close to the resonance
frequency and employing homodyne detection for monitoring the response. For
this detection scheme the normalized minimum detectable change in the
frequency $\sigma_{\mathrm{y}}$ is found to be given by \cite{Cleland_2758}%
\begin{equation}
\sigma_{\mathrm{y}}=\left(  \frac{2\Gamma k_{\mathrm{B}}T_{\mathrm{eff}}%
}{U_{0}\Omega^{2}\tau}\right)  ^{1/2}\ , \label{sigma_y ext}%
\end{equation}
where $k_{\mathrm{B}}$ is the Boltzmann's constant, $T_{\mathrm{eff}}$ is the
noise effective temperature and $U_{0}$ is the energy stored in the externally
driven resonator in steady state. Note that Eq. (\ref{sigma_y ext}) is derived
by assuming that the response of the resonator is linear and by assuming the
classical limit, i.e. $k_{\mathrm{B}}T_{\mathrm{eff}}\gg\hbar\Omega$. The
generalization of Eq. (\ref{sigma_y ext}) for the case of nonlinear response
is discussed in Ref. \cite{Buks_026217}.

In some cases applying external driving is technically difficult. For example,
for the case of the on-fiber optomechanical cavity, external driving that is
based on capacitive coupling requires electrical wiring to the device that is
located on the tip of the fiber as well as a closely positioned reference
electrode. On the other hand, these difficulties can be avoided by exploiting
the back-action effects resulting from optomechanical coupling, which can be
used for optically inducing self-excited oscillations of the mechanical
resonator.\cite{Zaitsev_046605} This method, which significantly simplifies
the operation since no oscillatory external drive is needed, can be employed
as an alternative resonant detection scheme.

In the limit of small displacement, the dynamics of the system can be
approximately described using a single evolution equation.\cite{Zaitsev_1589}
The theoretical model that is used to derive the evolution equation is briefly
described below. Note that some optomechanical effects that were taken into
account in the theoretical modeling \cite{Zaitsev_1589} were found
experimentally to have a negligible effect on the dynamics
\cite{Zaitsev_046605} (e.g. the effect of radiation pressure). In what follows
such effects are disregarded.

The micromechanical mirror in the optical cavity is treated as a mechanical
resonator with a single degree of freedom $x$ having mass $m$ and damping rate
$\gamma_{0}$ (when it is decoupled from the optical cavity). It is assumed
that the angular resonance frequency $\omega_{m}$ of the mechanical resonator
depends on the temperature $T$ of the suspended mirror. For small deviation of
$T$ from the base temperature $T_{0}$ (i.e. the temperature of the supporting
substrate) $\omega_{m}$ is taken to be given by $\omega_{m}=\omega_{0}%
-\beta\left(  T-T_{0}\right)  $, where $\beta$ is a constant. Furthermore, to
model the effect of thermal deformation \cite{Metzger_133903} it is assumed
that a temperature dependent force given by $F_{\text{\textrm{th}}}%
=\theta\left(  T-T_{0}\right)  $, where $\theta$ is a constant, acts on the
mechanical resonator. The mechanical oscillator's equation of motion is given
by $\ddot{x}+2\gamma_{0}\dot{x}+\omega_{m}^{2}x=F_{\text{\textrm{th}}}$, where
an overdot denotes differentiation with respect to time.

The time evolution of the effective temperature $T$ is governed by the thermal
balance equation $\dot{T}=\kappa\left(  T_{0}-T\right)  +\eta P_{\mathrm{L}%
}I\left(  x\right)  $, where $\eta$ is the heating coefficient due to optical
absorption, $\kappa$ is the thermal rate, $P_{\mathrm{L}}$ is the injected
laser power and $P_{\mathrm{L}}I\left(  x\right)  $ is the intra-cavity
optical power incident on the suspended mirror, which depends on the
mechanical displacement $x$ (i.e. on the length of the optical cavity) due to
the effect of optical interference. For small $x$ the expansion $I\left(
x\right)  \simeq I_{0}+I_{0}^{\prime}x+\left(  1/2\right)  I_{0}^{\prime
\prime}x^{2}$ is employed, where a prime denotes differentiation with respect
to the displacement $x$. This model neglects the mechanical nonlinearities of
the resonator, i.e. it is assumed that nonlinear behavior exclusively
originates from bolometric optomechanical coupling.

The displacement $x\left(  t\right)  $ can be expressed in terms of the
complex amplitude $A$ as $x\left(  t\right)  =x_{0}+2\operatorname{Re}A$,
where $x_{0}$, which is given by $x_{0}=\eta\theta P_{\mathrm{L}}I_{0}%
/\kappa\omega_{0}^{2}$, is the optically-induced static displacement. For
small displacement the evolution equation for the complex amplitude $A$ is
found to be given by \cite{Zaitsev_1589}%
\begin{equation}
\dot{A}+\left(  \Gamma_{\mathrm{eff}}+i\Omega_{\mathrm{eff}}\right)
A=\xi\left(  t\right)  \;,\label{A dot}%
\end{equation}
where both the effective resonance frequency $\Omega_{\mathrm{eff}}$ and the
effective damping rate $\Gamma_{\mathrm{eff}}$ are real even functions of
$\left\vert A\right\vert $. To second order in $\left\vert A\right\vert $ they
are given by%
\begin{align}
\Gamma_{\mathrm{eff}} &  =\gamma_{0}+\frac{\eta\theta P_{\mathrm{L}}}%
{2\omega_{0}^{2}}I_{0}^{\prime}+\frac{\eta\beta P_{\mathrm{L}}}{4\omega_{0}%
}I_{0}^{\prime\prime}\left\vert A\right\vert ^{2}\;,\label{Gamma_eff}\\
\Omega_{\mathrm{eff}} &  =\omega_{0}-\frac{\eta\beta P_{\mathrm{L}}}{\kappa
}I_{0}-\frac{\eta\beta P_{\mathrm{L}}}{\kappa}I_{0}^{\prime\prime}\left\vert
A\right\vert ^{2}\;.\label{Omega_eff}%
\end{align}
The above expressions for $\Gamma_{\mathrm{eff}}$ and $\Omega_{\mathrm{eff}}$
are obtained by making the following assumptions $\beta x_{0}\ll\theta
/2\omega_{0}$, $\theta\kappa^{2}\ll\beta\omega_{0}^{3}\lambda$ and $\kappa
\ll\omega_{0}$, which typically hold experimentally.\cite{Zaitsev_046605} The
fluctuating term \cite{Risken_Fokker-Planck} $\xi\left(  t\right)  =\xi
_{x}\left(  t\right)  +i\xi_{y}\left(  t\right)  $, where both $\xi_{x}$ and
$\xi_{y}$ are real, represents white noise and the following is assumed to
hold $\left\langle \xi_{x}\left(  t_{1}\right)  \xi_{x}\left(  t_{2}\right)
\right\rangle =\left\langle \xi_{y}\left(  t_{1}\right)  \xi_{y}\left(
t_{2}\right)  \right\rangle =2\Theta\delta\left(  t_{1}-t_{2}\right)  $ and
$\left\langle \xi_{x}\left(  t_{1}\right)  \xi_{y}\left(  t_{2}\right)
\right\rangle =0$ where $\Theta=\gamma_{0}k_{\mathrm{B}}T_{\mathrm{eff}%
}/4m\omega_{0}^{2}$ and where $T_{\mathrm{eff}}$ is the effective noise temperature.

In cylindrical coordinates $A$ is expressed as $A=A_{r}e^{iA_{\theta}}$, where
$A_{r}=\left\vert A\right\vert $ and where $A_{\theta}$ is
real.\cite{Hempstead_350} The equations of motion for $A_{r}$ and for
$A_{\theta}$ are given by [see Eq. (\ref{A dot})] $\dot{A}_{r}+A_{r}%
\Gamma_{\mathrm{eff}}=\xi_{r}\left(  t\right)  $ and $\dot{A}_{\theta}%
+\Omega_{\mathrm{eff}}\left(  A_{r}\right)  =\left(  1/A_{r}\right)
\xi_{\theta}\left(  t\right)  $, where the fluctuating terms satisfy the
following relations $\left\langle \xi_{r}\left(  t_{1}\right)  \xi_{r}\left(
t_{2}\right)  \right\rangle =\left\langle \xi_{\theta}\left(  t_{1}\right)
\xi_{\theta}\left(  t_{2}\right)  \right\rangle =2\Theta\delta\left(
t_{1}-t_{2}\right)  $ and $\left\langle \xi_{r}\left(  t_{1}\right)
\xi_{\theta}\left(  t_{2}\right)  \right\rangle =0$.

By introducing the notation $\Gamma_{0}=\gamma_{0}+\eta\theta P_{\mathrm{L}%
}I_{0}^{\prime}/2\omega_{0}^{2}$ and $\Gamma_{2}=\eta\beta P_{\mathrm{L}}%
I_{0}^{\prime\prime}/4\omega_{0}$ one can express the effective damping rate
$\Gamma_{\mathrm{eff}}$ as [see Eq. (\ref{Gamma_eff})] $\Gamma_{\mathrm{eff}%
}=\Gamma_{0}+\Gamma_{2}A_{r}^{2}$. Consider the case where $I_{0}^{\prime}<0$
and where $I_{0}^{\prime\prime}>0$. For such a case a supercritical Hopf
bifurcation occurs when $\Gamma_{0}$ vanishes, i.e. for a critical value of
the laser power given by $P_{\mathrm{LC}}=-2\omega_{0}^{2}\gamma_{0}%
/\eta\theta I_{0}^{\prime}$. The self-excited oscillations emerge above the
threshold, i.e. when $\Gamma_{0}$ becomes negative, with steady state value of
$A_{r}$ (when noise is disregarded) given by $r_{0}=\sqrt{-\Gamma_{0}%
/\Gamma_{2}}$, . In terms of the laser power $r_{0}$ can be expressed as%
\begin{equation}
r_{0}=\lambda\sqrt{-\frac{2\theta}{\beta\omega_{0}\lambda}\frac{I_{0}^{\prime
}}{\lambda I_{0}^{\prime\prime}}\frac{\Delta P_{\mathrm{L}}}{P_{\mathrm{L}}}%
}\;,\label{r_0}%
\end{equation}
where $\Delta P_{\mathrm{L}}=P_{\mathrm{L}}-P_{\mathrm{LC}}$.

Using the notation $A_{r}=r_{0}+\rho$ one finds to lowest nonvanishing order
in $\rho$ that $A_{r}\Gamma_{\mathrm{eff}}=-2\Gamma_{0}\rho+O\left(  \rho
^{2}\right)  $. Moreover $\Omega_{\mathrm{eff}}=\Omega_{\mathrm{H}}+\zeta
\rho+O\left(  \rho^{2}\right)  $, where $\Omega_{\mathrm{H}}=\Omega
_{\mathrm{eff}}\left(  r_{0}\right)  $ and where $\zeta=\mathrm{d}%
\Omega_{\mathrm{eff}}/\mathrm{d}A_{r}$ at the point $r_{0}$. To lowest
nonvanishing order in $\rho$ the equations of motion for $A_{r}$ and
$A_{\theta}$ become $\dot{\rho}-2\Gamma_{0}\rho=\xi_{r}\left(  t\right)  $ and
$\dot{\phi}+\zeta\rho=r_{0}^{-1}\xi_{\theta}\left(  t\right)  $, where
$\phi=A_{\theta}+\Omega_{\mathrm{H}}t$. In steady state (i.e. in the limit
$t\rightarrow\infty$) the solution for $\rho$ reads $\rho\left(  t\right)
=\int_{0}^{t}e^{2\Gamma_{0}\left(  t-t_{1}\right)  }\xi_{r}\left(
t_{1}\right)  \;\mathrm{d}t_{1}$. The correlation function of $\rho\left(
t\right)  $ is found to be given by $\left\langle \rho\left(  t_{1}\right)
\rho\left(  t_{2}\right)  \right\rangle =-\left(  \Theta/2\Gamma_{0}\right)
e^{2\Gamma_{0}\left\vert t_{1}-t_{2}\right\vert }$, and thus the correlation
function of $\dot{\phi}$ is given by%
\begin{equation}
\left\langle \dot{\phi}\left(  t_{1}\right)  \dot{\phi}\left(  t_{2}\right)
\right\rangle =\Theta\left[  \frac{\zeta^{2}e^{2\Gamma_{0}\left\vert
t_{1}-t_{2}\right\vert }}{2\left\vert \Gamma_{0}\right\vert }+2\frac
{\Gamma_{2}}{\left\vert \Gamma_{0}\right\vert }\delta\left(  t_{1}%
-t_{2}\right)  \right]  \;.
\end{equation}
With the help of the above result together with Wiener-Khinchine theorem one
finds that power spectrum $S_{\dot{\phi}}\left(  \omega\right)  $ of
$\dot{\phi}$ is given by%
\begin{equation}
S_{\dot{\phi}}\left(  \omega\right)  =\frac{1}{\pi}\frac{\Theta\zeta^{2}%
}{4\Gamma_{0}^{2}+\omega^{2}}+\frac{\Theta\Gamma_{2}}{\pi\left\vert \Gamma
_{0}\right\vert }\;.
\end{equation}

The signal $y\left(  t\right)  \equiv\dot{\phi}/\Omega_{\mathrm{H}}$
represents the normalized deviation of the momentary angular frequency
$\Omega+\dot{\phi}$ from its average value. The average value of\ $y\left(
t\right)  $ is estimated by monitoring the signal $y\left(  t\right)  $ in a
time interval $\tau$, i.e. $\hat{y}\left(  \tau\right)  =\tau^{-1}\int
_{-\tau/2}^{\tau/2}\mathrm{d}t\;y\left(  t\right)  $. In the limit of steady
state, i.e. when $\tau\gg1/\left\vert \Gamma_{0}\right\vert $, the variance of
the estimator $\hat{y}\left(  \tau\right)  $ is given by $\sigma_{\mathrm{y}%
}^{2}\left(  \tau\right)  =2\pi S_{\mathrm{y}}\left(  0\right)  /\tau$, thus
$\sigma_{\mathrm{y}}^{2}\left(  \tau\right)  =\left(  2\Theta/\Omega
_{\mathrm{H}}^{2}r_{0}^{2}\tau\right)  \left(  1+\zeta^{2}/4\left\vert
\Gamma_{0}\right\vert \Gamma_{2}\right)  $. Note that due to the fact that in
the present case $S_{\dot{\phi}}\left(  \omega\right)  $ remains finite in the
limit $\omega\rightarrow0$ the above defined variance $\sigma_{\mathrm{y}}%
^{2}\left(  \tau\right)  $ is identical to the Allan
variance.\cite{Allan_221,Cutler_136,Walls_162} With the help of Eq.
(\ref{Omega_eff}) one finds that $\zeta^{2}/4\left\vert \Gamma_{0}\right\vert
\Gamma_{2}=\left(  4\omega_{0}/\kappa\right)  ^{2}$. The assumption that
$\kappa\ll\omega_{0}$ \cite{Zaitsev_046605} leads to%
\begin{equation}
\sigma_{\mathrm{y}}\left(  \tau\right)  =\frac{4\omega_{0}}{\kappa}\left(
\frac{2\gamma_{0}k_{\mathrm{B}}T_{\mathrm{eff}}}{U_{0}\omega_{0}^{2}\tau
}\right)  ^{1/2}\;,\label{sigma_y SO}%
\end{equation}
where $U_{0}=4m\Omega_{\mathrm{H}}^{2}r_{0}^{2}$ is the energy stored in the
self-excited resonator in steady state close to the threshold. The comparison
with Eq. (\ref{sigma_y ext}) indicates that $\sigma_{\mathrm{y}}\left(
\tau\right)  $ for the case of optically induced self-excited oscillations is
roughly $4\omega_{0}/\kappa$ times larger compared with the case of external
drive for the same values of $U_{0}$ and $T_{\mathrm{eff}}$.

\begin{figure}
[ptb]
\begin{center}
\includegraphics[
height=2.4396in,
width=3.2396in
]%
{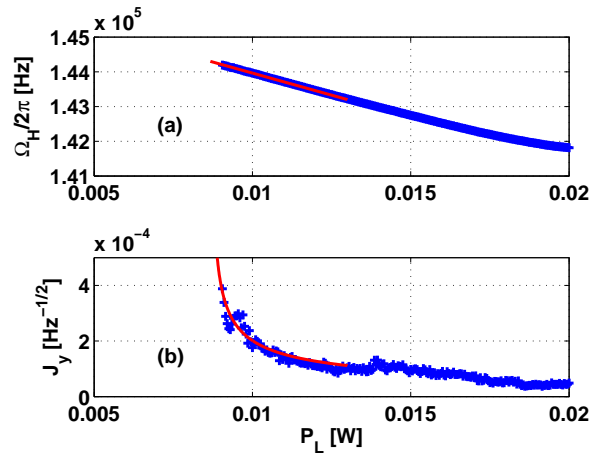}%
\caption{The frequency of self-excited oscillations $\Omega_{\mathrm{H}}/2\pi$
[panel (a)] and the sensitivity factor $J_{\mathrm{y}}$ [panel (b)] vs. laser
power $P_{\mathrm{L}}$. The blue cross marks represent experimental data, the
red solid lines represent a theoretical fit that is based on the following
experimental parameters: $\kappa=22\times10^{3}\operatorname{Hz}$,
$m=30\times10^{-12}\operatorname{kg}$, $\lambda=1.55\operatorname{\mu m}$,
$T_{\mathrm{eff}}=300\operatorname{K}$, $\Omega/\Gamma=10^{3}$, $\left\vert
I_{0}^{\prime}\right\vert /\lambda I_{0}^{\prime\prime}=0.8$, $\beta
=10^{4}\operatorname{s}^{-1}\operatorname{K}^{-1}$, $\theta=1.6\times
10^{-3}\operatorname{N}\operatorname{kg}^{-1}\operatorname{K}^{-1}$ and
$\eta=3.5\times10^{6}\operatorname{K}\operatorname{J}^{-1}$.}%
\label{SO_data_fit}%
\end{center}
\end{figure}

In general, the sensitivity factor $J_{\mathrm{p}}$ of a physical parameter
$p$ is given by $J_{\mathrm{p}}=\left\vert \partial\log\Omega/\partial
p\right\vert ^{-1}J_{\mathrm{y}}$, where $J_{\mathrm{y}}=\sigma_{\mathrm{y}%
}\tau^{1/2}$ represents the sensitivity factor of the parameter $y$. The
factor $J_{\mathrm{y}}$ as function of $P_{\mathrm{L}}$\ is experimentally
measured using an on-fiber optomechanical cavity device, in which the
mechanical mirror covers almost the entire fiber cross section. Self-excited
oscillation as a function of input power has been observed above a threshold
value of $P_{\mathrm{LC}}=8.7%
\operatorname{mW}%
$ [see Fig. \ref{SO_data_fit} (a)]. To experimentally determine the factor
$J_{\mathrm{y}}$ the reflected optical power is recorded over a time period of
$2%
\operatorname{ms}%
$ using an oscilloscope. The standard deviation $\sigma_{\mathrm{y}}$, which
is found by the zero-crossing technique \cite{Cutler_136}, allows evaluation
of the factor $J_{\mathrm{y}}=\sigma_{\mathrm{y}}\tau^{1/2}$. The frequency of
self-excited oscillations $\Omega_{\mathrm{H}}/2\pi$ and the sensitivity
factor $J_{\mathrm{y}}$ are plotted vs. laser power $P_{\mathrm{L}}$ in panels
(a) and (b) respectively of Fig. \ref{SO_data_fit}. The red solid line in
panel (a) represents a theoretical fit that is based on Eq. (\ref{Omega_eff})
and the one in panel (b) is based on Eqs. (\ref{r_0}) and (\ref{sigma_y SO}).
The experimental parameters that have been employed in both cases are listed
in the figure caption. Similar measurements of the factor $J_{\mathrm{y}}$
with other on-fiber devices (e.g. results shown in the inset of Fig.
\ref{Fig_osc}) as well as with devices that were made on a wafer have yielded
values having the same order of magnitude [as in the data seen in Fig.
\ref{SO_data_fit}(b)].%

In summary, optically induced self-excited oscillations of suspended mirror
fabricated on tip of an optical fiber were experimentally demonstrated. Based
on previously developed model of bolometrically coupled optomechanical cavity
we estimated the sensitivity of a detector exploiting the self-excited
oscillations effect. The fabrication and operation simplicity of self-excited
oscillating optomechanical detectors in many cases compensates the possible
degradation in sensitivity compared to traditional externally actuated
resonator. 

This work was supported by the German Israel Foundation under Grant No.
1-2038.1114.07, the Israel Science Foundation under Grant No. 1380021, the
Deborah Foundation, the Mitchel Foundation, the Israel Ministry of Science,
the Russell Berrie Nanotechnology Institute, the European STREP QNEMS Project,
MAGNET Metro 450 consortium and MAFAT. The work of IB was supported by a Zeff Fellowship.

\newpage

\bibliographystyle{ieee}
\bibliography{acompat,Eyal_Bib}

\end{document}